\documentclass[sigconf]{acmart}
\AtBeginDocument{%
  }
\usepackage{mathrsfs}
\usepackage{amsthm}
\usepackage{dsfont}

\newtheorem{lemma}{Lemma}
\newtheorem{definition}{Definition}
\newtheorem{claim}{Claim}
\usepackage{algorithm}
\usepackage{algorithmic}
\usepackage{tcolorbox} 
\usepackage{varwidth} 
\tcbuselibrary{breakable} 
\tcbuselibrary{skins} 
\NewTColorBox{theobox}{o m +o}{
    enhanced,
    colframe=gray!0!white,
    interior empty,
    coltitle=white,
    fonttitle=\mdseries,
    colbacktitle=gray,
    extras broken={frame empty,interior empty},
    borderline={0.5mm}{0mm}{gray},
    sharp corners=downhill,
    top=4mm,
    before skip=3.5mm,
    attach boxed title to top left={yshift=-3mm,xshift=5mm},
    boxed title style={boxrule=0pt,sharp corners=all},varwidth boxed title,
    IfNoValueTF={#1}{title=#2}{title=#2:~#1},
    IfNoValueTF={#3}{}{#3}
}
\setcopyright{acmlicensed}
\copyrightyear{2025}
\acmYear{2025}

\acmConference[ArXiv '2501]{Preprint}{}{}
\acmBooktitle{Archive Preprint}
\acmISBN{978-1-4503-XXXX-X/18/06}

\begin{document}

\title{Let Watermarks Speak: A Robust and Unforgeable Watermark for Language Models}

\author{Minhao Bai}
\email{bmh22@mails.tsinghua.edu.cn}
\affiliation{%
  \institution{Tsinghua University}
  \city{Beijing}
  \country{China}
}









\begin{abstract}
    Watermarking is an effective way to trace model-generated content. Current watermark methods cannot resist forgery attacks, such as a deceptive claim that the model-generated content is a response to a fabricated prompt. None of them can be made unforgeable without degrading robustness. 

    Unforgeability demands that the watermarked output is not only detectable but also verifiable for integrity, indicating whether it has been modified. This underscores the necessity and significance of a multi-bit watermarking scheme.
    Recent works try to build multi-bit scheme based on existing zero-bit watermarking scheme, but they either degrades the robustness or brings a significant computational burden. We aim to design a novel single-bit watermark scheme, which provides the ability to embed 2 different watermark signals. 
    

    This paper's main contribution is that we are the first to propose an undetectable, robust, single-bit watermarking scheme. It has a comparable robustness to the most advanced zero-bit watermarking schemes.  Then we construct a multi-bit watermarking scheme to use the hash value of prompt or the newest generated content as the watermark signals, and embed them into the following content, which guarantees the unforgeability. 

    Additionally, we provide sufficient experiments on some popular language models, while the other advanced methods with provable guarantees do not often provide. The results show that our method is practically effective and robust.
\end{abstract}

\ccsdesc[500]{Security and privacy~Security services}
\ccsdesc[300]{Security and privacy~Cryptography}
\ccsdesc[100]{Security and privacy~Network security}

\keywords{Watermarking, Generative Models, IP Protection, Applied Cryptography}


\maketitle

\section{Introduction}
Language models \cite{} have undergone significant development over the past five years. Developers have gone to great lengths to evolve the performance of the models to a point where they are close to humans, but the model-generated content may be used in illegal or harmful ways. 
Since the model-generated content is a type of virtual product, it should be well labeled as natural products, which can help counteract the misuse of them.
To counteract the misuse of the model-generated content, one of the most effective approach is watermarking.

Watermarking is an active methods that embedding a signal into the model-generated contents rather than passively distinguishing between them and normal texts. Current watermarking methods modify the output distribution of model or sample from the distribution as they want. No matter how they are constructed, most of them are robust, which means that certain modification on the watermarked output does not affect the detection of watermarks. 
As one of the most robust watermarking methods, \cite{kirchenbauer2023watermark} suggests that a long enough segment of watermarked output should be still detectable. Especially, the distortion-free watermarking methods \cite{christ2024pseudorandom,Aar,cohen2024watermarking,kuditipudi2024robust} do not affect the quality of models' output while maintains the same level robustness as \cite{kirchenbauer2023watermark}. 

However, as most of the researchers explore the robustness of watermark, we notice that some simple forgery attacks are possible if the watermarked output is robust. As these robust watermarking methods generate robust watermarked text, simply modifying a few words while controlling that the watermark signal can still be detected is easy. Though the modified text is not directly generated by the model and its meaning may change a lot, it will trouble the model's provider if the modified text is offensive and detectable. For most of the current watermarking techniques, the robustness and the unforgeability are contradict. We name such attacks as \textbf{robustness-exploiting forgery attacks}.

A more problematic scenario is that the attacker put the watermarked text under the wrong prompt, without any modification. For example, we let ChatGPT answer the following 2 questions: (1) \textit{What is the best way to cook a steak?} and (2) \textit{What should I do if my pet passed away?} And the ChatGPT outputs 2 answers: (1) \textit{Sear it in a hot pan and finish it in the oven for the perfect medium-rare.} and (2) \textit{Consider contacting a pet cremation service or discussing options with your veterinarian.} If we claim the ChatGPT gets question (2) and outputs the answer (1), it will be a successful frame if answer (1) is detected as watermarked. Current watermarking methods are unable to clear themselves in such scenario. We name such attacks as \textbf{prompt-misattribution forgery attacks}.

These forgery attacks show some blind spots of existing watermarking methods. To avoid the robustness being used for forgery, the detector should be aware that whether the watermarked text has been modified. A feasible solution is using the hash function or a signature scheme to verify its integrity. To avoid the prompt-misattribution forgery attacks, the watermark signals that embedded into the text should represent some information about the real prompt. Both of them require  a multi-bit watermarking scheme. 

However, current distortion-free watermarks may loss some robustness in the way to unforgeablility. 
\cite{christ2024pseudorandom} directly extents a zero-bit scheme into a multi-bit scheme by using watermarked output as signal 1 and unwatermarked output as 0, which will significantly weaken the original robustness. For an instance, embedding a watermark signal 0 results in some unwatermarked output. Such unwatermarked output can escape from detector though they are generated by the model. \cite{fairoze2024publiclydetectablewatermarkinglanguagemodels} claims an unforgeable watermarking scheme under the assumption that the model generated text has a minimum entropy, which is not realistic in the current language models \cite{METEOR}. This work is based on a classic steganography method \cite{hopper} and only has a weak robustness. In conclusion, none of the previous methods successfully achieves unforgeability without loss of robustness.

Motivated by the contradiction between robustness and unforgeability, we aim at proposing an unforgeable watermarking scheme without degrading robustness. 
The main contribution and novelty of this paper is that our construction begins from a natural single-bit scheme but not a zero-bit scheme. The proposed single-bit scheme is able to embed 2 different watermark signals. The single-bit detector can distinguish between 3 states: 0, 1 and $\perp$, the notion of no signal, with a robustness comparable to \cite{undetect}. For the non-watermarked text, the detector only outputs $\perp$. But for watermarked output, the detector outputs a bit 0 or 1. It naturally follows a multi-bit scheme.

To prevent the attacks that attribute the right answer to the wrong prompt, We hash the prompt and embed each bit of the hash into the following output. To identify that whether the output has been modified, we hash the newest generated output and embed each bit of the hash into the next part of output. After some modification on the watermarked output, the hash bits can still be decoded due to the robustness, but the hash of prompt or previous watermarked text is not identical to the embedded hash bits.  Even if the watermarked text is severely modified, it is still detectable as long as the detector find a bit. Therefore, our construction can remind the detector of the condition of watermarked text and help to locate the modification, which guarantees the unforgeability.

Main contributions of this paper are summarized as follows:
\begin{itemize}
    \item We propose a single-bit watermarking scheme, which naturally contains 3 states and 2 of them are symmetric.
    \item We propose a multi-bit watermarking scheme, which is prefix-unforgeable and robust.
    \item The proposed watermarking schemes generate text that is computationally indistinguishable from the model generated text, and do not explicitly decrease the generation speed.
\end{itemize}

\section{Relate Work}
\subsection{Text Classifiers}

Before the dawn of practical language models, most of the watermark works \cite{seqxgpt,radar,detectgpt,fastdetectgpt,yang2023zeroshotdetectionmachinegeneratedcodes,binocular} focus on the natural features of model-generated text. They build classifiers based on these features and deep learning modules are widely used. They can be roughly divided into 2 classes: the training methods \cite{seqxgpt,radar} and trainless methods. Training methods involves collecting a dataset that includes natural text and model-generated text, and then capturing the feature difference by training a well-designed deep learning module, such as convolution networks with self-attention \cite{seqxgpt}, adversarial network \cite{radar} , and so on. This class of methods is possibly limited and biased by the provided data, as the diversity of text is far beyond the ability of these deep learning modules. Trainless methods \cite{detectgpt,fastdetectgpt,yang2023zeroshotdetectionmachinegeneratedcodes,binocular} rely on some explicit features, such as the probability curvature \cite{detectgpt,fastdetectgpt,yang2023zeroshotdetectionmachinegeneratedcodes} and different views of models \cite{binocular}. However, without the help of data, trainless methods empirically have an unstable performance. 

With the development of language models, the model-generated text becomes more and more similar to natural text. Unstable and unexplainable classifiers may not be a suitable solution of detecting model-generated text.
\subsection{Zero-bit Watermarks}

As the text classifiers do not perform well in the current age, most of the attention has been paid to methods that actively inject a specified signal. Current zero-bit watermarking schemes can be classified to 2 groups: distorted watermarks and distortion-free watermarks. As the pioneer of distorted watermarks, Red-Green watermark \cite{kirchenbauer2023watermark} shows an empirical way to build robust watermark by partitioning the vocabulary into green list and red list. It makes the probability of outputting a token in green list increased, leading to a majority of output tokens are in green list. This work proposes the basic properties that a watermark should have: substring detectable without the model and robust to a certain degree of modification. However, as the Red-Green watermark modifies the output distribution of the model, the watermarked text may suffer from degradation. Based on this method, a group of distorted but robust watermarking methods \cite{zhao2024provable,kirchenbauer2024reliabilitywatermarkslargelanguage,liu2024an} are proposed. These works establish the framework of robust watermarks, and their negative impact on text quality motivates a series of distortion-free watermarks.

In fact, the research of distortion-free watermarks may even predate the Red-Green watermark. Early in 2023, \cite{Aar} has proposed a watermark based on gumble-softmax sampling \cite{jang2017categorical}. As the token output by gumble-softmax sampling obeys the distribution of original language model, it does little harm to the text quality. However, as the randomness used for gumble-softmax sampling is derived from a pseudorandom generator with preceding tokens as input, the preceding tokens should not repeat to keep the output of pseudorandom generator is really pseudorandom. \cite{undetect} builds an undetectable watermarking scheme based on \cite{Aar}. This work uses preceding tokens with enough entropy to feed into the pseudorandom generator, which avoids possible repetitive inputs. 
\cite{christ2024pseudorandom} proposes an updated version of \cite{undetect}, which is robust against deletion. Another robust and distortion-free watermark method \cite{golowich2024editdistancerobustwatermarks}  follows \cite{christ2024pseudorandom} but relies on weaker assumptions. 

These watermark methods focus on the robustness and text quality, and few of them mention the problem of forgery. The class of Red-Green watermark are forgeable, as their output text can be modified without removing the watermark, which is guaranteed by their robustness. For the class of distortion-free watermarks, they cannot prevent misattribution forgery attacks as they do not protect the prompt that used to generate watermarked text.

\subsection{Multi-bit Watermarks}

In order to provide more details of the model-generated text rather than a single signal, multi-bit watermarks \cite{qu2024provablyrobustmultibitwatermarking,cohen2024watermarking,fairoze2024publiclydetectablewatermarkinglanguagemodels,christ2024pseudorandom,zamir2024excuse} are necessary. These methods can still be partitioned into Red-Green based \cite{qu2024provablyrobustmultibitwatermarking} and distortion-free based \cite{cohen2024watermarking,fairoze2024publiclydetectablewatermarkinglanguagemodels,christ2024pseudorandom,zamir2024excuse}. \cite{christ2024pseudorandom} also provides a multi-bit construction, by letting watermarked output as 1 and uniformly sampled output as 0. However, such straight forward expansion may degrade its robustness. As uniformly sampled output is the same as unwatermarked output, some substring may not be correctly detected. \cite{cohen2024watermarking} follows the zero-bit scheme of \cite{christ2024pseudorandom}, and extends to multi-bit by sampling multiple keys. Though such construction still maintains the robustness and undetectability of \cite{christ2024pseudorandom}, this method suits cases with fewer watermark signals to embed. Encoding $L$-bit requires $2^L$ times of executing zero-bit detector, which is not affordable in our scenario. 

In the multi-bit watermarks, only \cite{fairoze2024publiclydetectablewatermarkinglanguagemodels} focuses on the forgery problem. This method utilizes digital signature scheme based on a steganography construction of \cite{hopper}. As this method can also extract bits from unwatermarked text, one should wait until the extraction finish and check the signature. The steganography construction of \cite{hopper} requires the assumption of minimum entropy, which indicates that $k$ tokens should provide more than $\alpha$ bits entropy. However, this assumption has been proved not suitable for language models \cite{METEOR}.

This paper follows \cite{fairoze2024publiclydetectablewatermarkinglanguagemodels}, \cite{christ2024pseudorandom} and \cite{cohen2024watermarking} to build a multi-bit watermark. Our method begins from a natural single-bit construction rather than relies on an existed zero-bit construction, which is the main difference in the mechanism.
\section{Preliminaries}

\subsection{Notations}
We denote the polynomial function as $\textsf{poly}(n)$, which has an equivalent order to the function $n^c$, where $c > 0$ is a constant. We denote the negligible function as $\textsf{negl}(n)$, which is asymptotically smaller than any inverse of polynomial function. We use $\lambda$ to denote the secure parameter, which indicates the degree of security. 

In this paper, $\{0,1\}^*$ denotes the set of all bit strings with finite length. For a bit string $b$, we use $b_{i}$ to denote the $i$-th bit of $b$, and we use $b_{i:j}$ to denote a substring that begins from the $i$-th bit and ends at the $j$-th bit. The concatenation of 2 strings $a, b$ is denoted as $a || b$. We use the symbol $\emptyset$ to denote an empty string. The length of a string $b$ is denoted by $|b|$. The Hamming weight of a string $b$ is denoted by $\mathsf{wt}(b)$, which represents the number of ones in $b$.

We use \textsf{Unif} to denote the uniform distribution. For example, $\mathsf{Unif}[0,1]$ represents a random number between 0 and 1. We use $x \xleftarrow{\$} \mathsf{Unif}[0,1]$ to denote sampling $x$ from the distribution $\mathsf{Unif}[0,1]$.
\subsection{Model the Auto-regressive Model}
We define an auto-regressive model, \textsf{Model}, as a pair of deterministic algorithm \textsf{Predictor} and a probabilistic algorithm \textsf{Sampler}, $\textsf{Model} = (\textsf{Predictor}, \textsf{Sampler})$.
\begin{itemize}
    \item $\textsf{Predictor}(z)$ takes arbitrary input $z \in \{0,1\}^*$ and output a distribution $D_z$ of the next bit.
    \item $\textsf{Sampler}(D_z,r)$ takes the distribution $D_z$ as input and output a single bit $b \in \{0,1\}$ guided by some randomness $r$.
\end{itemize}
The \textsf{Model} can be seen as a model that takes bits as input and outputs a bit. Though current language models does not only output bits but tokens, converting these models into \textsf{Model} is not difficult. For an instance, the token ID can be translated into a list of bits. We can first decide to sample from the tokens whose bit lists start with 0 or 1. This sampling procedure exactly generates a bit. If we first sample from the tokens whose bit lists start with 0, we can then decide to sample from the tokens whose bit lists start with 0,0 or 0,1. In this way, a single generation step of the language models can be converted into several generation steps of \textsf{Model}.

\subsection{Pseudorandom Generator}
A deterministic algorithm $\textsf{PRG}(k)$ is a pseudorandom generator if for all polynomial time adversaries $\mathcal{A}$, 
\begin{align}
   \left|\mathbf{Pr}\left[\mathcal{A}^{\textsf{PRG}(k)} = 1\right] - \mathbf{Pr}\left[\mathcal{A}^{U\sim\{0,1\}^{\mathsf{poly}(\left|k\right|)}} = 1\right] \right| < \mathsf{negl}(\left|k\right|).
\end{align}
Once the $\textsf{PRG}(k)$ generates a list of $\mathsf{poly}(\left|k\right|)$ random bits $\{rb_1, rb_2, \cdots\}$, they can be converted into a real number in [0,1] by taking $\sum_{i=1}^{\mathsf{poly}(\left|k\right|)} \frac{rb_i}{2^i}$. In the following, we always convert the output of $\textsf{PRG}(k)$ into these real numbers and write them as $\{r_1, r_2, \cdots\}$.
\subsection{Collision Resistance Hash Function}
A deterministic, polynomial time computable function $\textsf{Hash} \{0,1\}^n \xrightarrow{} \{0,1\}^m, n>m$ is a collision resistance hash function if for all polynomial time adversary $\mathcal{A}$, 
\begin{align}
    \mathbf{Pr}\left[ \mathcal{A}^\textsf{Hash}(\cdot)\longrightarrow (x_n,y_n), \textsf{Hash}(x_n) = \textsf{Hash}(y_n) \right] < \mathsf{negl}(n).
\end{align}
\subsection{Zero-bit Watermarking Scheme}
A zero-bit watermarking scheme $\textsf{Wat}$ is defined as a triple of possibly probabilistic algorithms, $\textsf{Wat} = (\textsf{KeyGen}, \textsf{Embed}, \textsf{Detect})$. 
\begin{itemize}
    \item $\mathsf{KeyGen}(1^\lambda)$ takes a random bit string of length $\lambda$ as input and output a key $k \in \{0,1\}^{\textsf{poly}(\lambda)}$.
    \item $\mathsf{Embed}(k,m,z)$ takes a key $k$, watermark signal $m \in \{1\}$, and \textsf{Model}'s input $z \in \{0,1\}^*$ as input. It runs $\textsf{PRG}$ and $\textsf{Model}$ to output a series of bit $b\in\{0,1\}^{\textsf{poly}(\lambda)}$.
    \item $\mathsf{Detect}(k,b)$ takes a key $k$ and a series of bit $b\in\{0,1\}^{\textsf{poly}(\lambda)}$ as input. It outputs a signal $m \in \{1,\bot\}$.
\end{itemize}



\subsection{Multi-bit Watermarking Scheme}
The main difference between the multi-bit scheme and the zero-bit scheme is the set of watermark signal is larger than 1.  Its $\mathsf{Embed}(k,m,z)$ algorithm takes a key $k$, watermark signal $m \in \{0,1\}^{\mathsf{poly}(\lambda)}$, and \textsf{Model}'s input $z \in \{0,1\}^*$ as input. And its $\mathsf{Detect}(k,b)$ algorithm takes a key k and a series of bit $b\in\{0,1\}^{\mathsf{poly}(\lambda)}$ as input. It outputs a signal $m \in \{0,1,\bot\}^{\mathsf{poly}(\lambda)}$.

A multi-bit watermarking scheme is \textbf{correct} if for $\forall k \in \{0,1\}^{\textsf{poly}(\lambda)}, \forall m \in \{0,1\}^{\mathsf{poly}(\lambda)}, \forall z \in \{0,1\}^*$,
\begin{align}
    \mathbf{Pr}\left[\mathsf{Detect}(k,\mathsf{Embed}(k,m,z)) = m\right] > 1-\mathsf{negl}(\lambda)
\end{align}
and if for any $b \not\in \{\mathsf{Embed}(k,m,z)): \forall k \in \{0,1\}^{\textsf{poly}(\lambda)}, \forall z \in \{0,1\}^* \}$,
\begin{align}
    \mathbf{Pr}\left[\mathsf{Detect}(k,b) = \{\bot\}\right] > 1-\mathsf{negl}(\lambda).
\end{align}


It is worthy to mention that convert a zero-bit watermarking scheme into a multi-bit watermarking scheme is possible by running $\mathsf{Embed}$ as signal 1 and running $\mathsf{Model}$ with random $r$ as signal 0 \cite{christ2024pseudorandom}. And if at the end of text the $\mathsf{Detect}$ only outputs 0, it is unwatermarked with high probability. However, such straight forward expansion may yield problem if the embedded watermark signal is always 0, which results that the watermark algorithm generates an unwatermarked text. Though the above condition can be avoided by not allowing the watermark signal is full of 0, their $\mathsf{Detect}$ algorithm is significantly weakened because unwatermarked content exists. Especially, if the watermark is publicly detectable, those unwatermarked part may become the target of modification.

Another way is to use several different keys to represent different messages \cite{cohen2024watermarking}. It may cause an unaffordable computation burden if the message is long.



\section{A Robust and Unforgeable Watermark}

\subsection{A Single-bit Watermarking Scheme}

Unlike those works that are built from a zero-bit watermarking scheme, we straightly begin with a single-bit watermarking scheme $\mathcal{W} = (\mathsf{KeyGen},$ $ \mathsf{Embed}_{sk}, \mathsf{Detect}_{sk})$. 


\begin{definition}[Single-bit Watermarking Scheme]
A single-bit watermarking scheme for a language model \textsf{Model} is a tuple of efficient algorithms $\mathcal{W} = (\textsf{KeyGen}, \textsf{Embed}, \textsf{Detect})$ where:
\begin{itemize}
    \item $\textsf{KeyGen}(1^\lambda) \xrightarrow{} sk$ is a randomized algorithm that takes the security parameter $\lambda$ as input, and outputs a secret key $sk$.
    \item $\textsf{Embed}_{sk}(z,m)\xrightarrow{} \mathbf{b}$ is a keyed randomized algorithm that takes prompt $z$ and watermark signal $m$ as input, and outputs a block of watermarked bits $\mathbf{b}$.
    \item $\textsf{Detect}_{sk}(\mathbf{b'}) \xrightarrow{} (m', \textsf{block})$ is a keyed randomized algorithm that takes a bit string $\mathbf{b'} \in \{0,1\}^*$ as input, and outputs a watermark signal $m' \in \{0,1,\bot\}$. If $m' \neq \bot$, it output a substring \textsf{block} of $\mathbf{b'}$ that contains the watermark signal $m'$. Otherwise it output an empty string.
\end{itemize}
\end{definition}

Before we construct \textsf{Embed} and \textsf{Detect}, we mentioned that in the definition of \textsf{Model}, it does not take a watermark signal as input. We focus on the \textsf{Sampler} of \textsf{Model}, and substitutes it with \textsf{Wat-Sampler}, which takes the watermark signal as an additional input. We call the pair (\textsf{Predictor},  \textsf{Wat-Sampler}) as \textsf{Wat-Model}. Symmetrically, we construct an algorithm \textsf{Detect-1bit} to provide the confidence on each bit generated by \textsf{Wat-Sampler}.

\textbf{Construction of \textsf{Wat-Sampler}}.
We denote that the probability that the next bit is 0 as $p(0)$ and the probability that the next bit is 1 as $p(1)$. $\textsf{Wat-Sampler}(r,(p(0),p(1)), m)\xrightarrow{} b$ takes a watermark signal $m \in \{0,1\}$, the distribution $p(0),p(1)$ and a random number $r \in [0,1]$ as input, and output a single text bit $b$ by the following computation:

\begin{align}
    b = \mathds{1}(r < p(b)) \oplus m
\end{align}
\begin{figure}
    \centering
    \includegraphics[width=\linewidth]{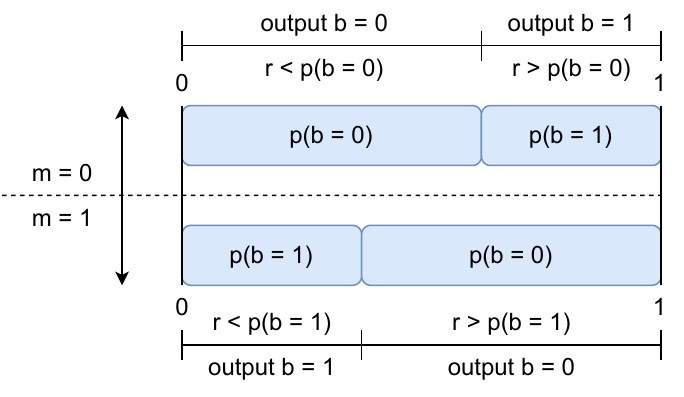}
    \caption{An illustration of \textsf{Wat-Sampler} and DITS.}
    \label{Wat-sampler}
\end{figure}
We call the above sampling method as Dual Inverse Transform Sampling (DITS), and details are shown in Fig. \ref{Wat-sampler}. \textsf{Wat-Sampler} actually constructs 2 dual copies of distribution with different arrangement, and it selects one of the two copies based on the value of $m$ to do inverse transform sampling. 

The output of \textsf{Wat-Sampler} follows the distribution $(p(0), p(1))$ regardless of the distribution of watermark signal $m$ and we formally prove this in the follows.

\begin{claim}[DITS Maintains the Distribution of \textsf{Predictor}]
    For any distribution $(p(0),p(1))$ outputted by \textsf{Predictor}, the distribution of DITS output bit $b$ obeys the original distribution $(p(0),p(1))$ regardless of the distribution of watermark signal $m \in \{0,1\}$.
\end{claim}
\textit{Proof.}  Recall that the random number $r \sim \textsf{Unif}[0,1]$. As shown in Fig. \ref{Wat-sampler}, the probability of $b = 0$ is:
\begin{align}
    \mathbf{Pr}\left[ b = 0 \right] & = \mathbf{Pr}\left[ r < p(0) \right] \cdot \mathbf{Pr}\left[ m = 0 \right] + \mathbf{Pr}\left[ r > p(1) \right] \cdot \mathbf{Pr}\left[ m = 1 \right] \notag \\
    & = p(0)\mathbf{Pr}\left[ m = 0 \right] + (1 - p(1))\mathbf{Pr}\left[ m = 1 \right] \notag \\ 
    & = p(0) \left(\mathbf{Pr}\left[ m = 0 \right] + \mathbf{Pr}\left[ m = 1 \right]\right) \notag \\ 
    & = p(0),
\end{align}
which is independent from the watermark signal $m$.
Similarly, the probability of $b = 1$ is identical to $p(1)$, regardless of the distribution of $m$. 

\textbf{Construction of \textsf{Detect-1bit}}.
The algorithm $\textsf{Detect-1bit}(r,b)\xrightarrow{} m'$ takes the same random number $r$ and the bit $b$ outputted by \textsf{Wat-Sampler} as input, and outputs a watermark signal $m'$. Based on the behavior of \textsf{Wat-Sampler}, it works as follows:
\begin{align}
    m' = \mathds{1}(r < 1/2) \oplus b.
\end{align}
\begin{figure}
    \centering
    \includegraphics[width=\linewidth]{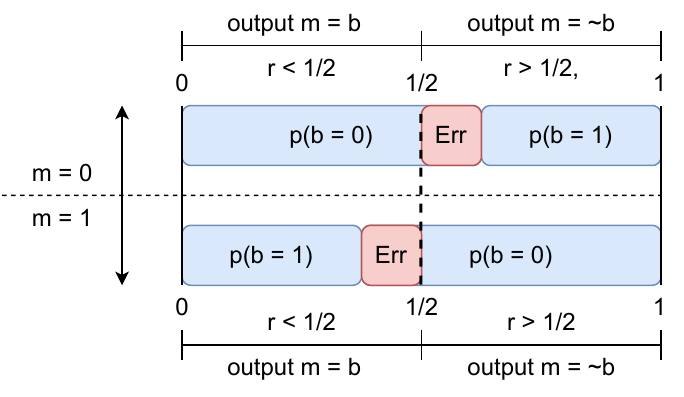}
    \caption{An illustration of \textsf{Detect-1bit}.}
    \label{fig:enter-label}
\end{figure}
So in the case that $p(0)=p(1)=1/2$, the \textsf{Detect-1bit} always outputs $m' = m$. If $p(0) > 1/2$ and the $\textsf{Wat-Sampler}$ takes $m=0$ as input, the probability of $r > 1/2$ and output a bit $b=0$ is $p(0) - 1/2$. As shown in Fig. \ref{detect1bit}, the \textsf{Detect-1bit} will output a wrong watermark signal $m'=1$.  Similarly, if $p(1) > 1/2$ the probability of \textsf{Detect-1bit} outputs a wrong watermark signal is $p(1) - 1/2$. Because $p(0) - 1/2 < 1/2$ and $p(1) - 1/2 < 1/2$, \textsf{Detect-1bit} has an error probability of at most $1/2$.

\textbf{Aggregate the output of \textsf{Detect-1bit} on multiple bits}.
\textsf{Wat-Sampler} can be executed several steps by $\mathsf{Embed}_{sk}$ to repeatedly embed the same watermark signal $m$. As the error probability is at most $1/2$ but not always $1/2$, \textsf{Detect-1bit} should output more correct watermark signals.
Thus, aggregating the output of \textsf{Detect-1bit} on multiple bits and making a decision bound are necessary. Assuming that $\mathsf{Embed}$ generates a watermarked bit string $b$, the distribution of each bit is $(p_i(0), p_i(1)), i \in \{1,2,\cdots, |b|\}$ and the randomness used to sample each bit is $r_i$. For the $i$-th bit $b_i$, the probability that \textsf{Detect-1bit} outputs a wrong watermark signal is $\max\{p_i(0),p_i(1)\} - 1/2$. The expectation of \textsf{Detect-1bit}'s output when the real watermark signal $m=0$ is 
\begin{align}
    &\mathbb{E}\left[\textsf{Detect-1bit}(b_i,r_i) \mid b_i = \textsf{Wat-Sampler}(r_i, (p_i(0),p_i(1)), m =0)\right] \notag\\
    = &0 \cdot \left[ 1- (\max\{p_i(0),p_i(1)\} - 1/2)\right] + 1 \cdot (\max\{p_i(0),p_i(1)\} - 1/2) \notag\\
    = &\max\{p_i(0),p_i(1)\} - 1/2.
\end{align}
Similarly, The expectation of \textsf{Detect-1bit}'s output when the real watermark signal $m=1$ is $3/2 - \max\{p_i(0),p_i(1)\}$. As these 2 conditions lead to different expectations, \textsf{Detect} is able to distinguish them within sufficient number of watermarked bits.
\begin{lemma}[Hoeffding's Inequality]\label{Ho}
$X_1, X_2, \cdot\cdot\cdot , X_n$ are independent and identical random variables, and each $X_i$ is bounded by $[0,1]$.
Let $X = \frac{1}{n}\sum_{i=1}^n X_i$ and $\mu = \mathbb{E}[X]$, the probability that the sample mean $X$ deviates from the theoretical mean $\mu$ up to $t$ is 
\begin{align}
    \mathbf{Pr}\left[X - \mu \geq t\right] \leq \exp\left(-2{nt^2}\right),
\end{align}
and 
\begin{align}
    \mathbf{Pr}\left[\mu - X \geq t\right] \leq \exp\left(-2{nt^2}\right).
\end{align}
\end{lemma}
We denote the outputs of \textsf{Detect-1bit} as a series of random variables $X_1, X_2, \cdots, X_N$, $X_i = \mathds{1}(r_i < 1/2) \oplus b_i$. The algorithm \textsf{Detect} aggregates them to compute $X = \frac{1}{N}\sum_{i=1}^N X_i$. 
For all of the $N$ watermarked bits embedded with identical watermark signals, the conditional expectation of $X$ is 
\begin{align}
    \mathbb{E}\left[X \mid b = \mathsf{Embed}_{sk}(z,m=0)\right]  & = \frac{1}{N}\left(\sum_{i=1}^N \max\{p_i(0),p_i(1)\} \right) - 1/2 \notag\\
    & = 1/2 - \frac{1}{N}\left(\sum_{i=1}^N \min\{p_i(0),p_i(1)\}\right)
\end{align}
and 
\begin{align}
    \mathbb{E}\left[X \mid b = \mathsf{Embed}_{sk}(z,m=1)\right]  & = 3/2 - \frac{1}{N}\left(\sum_{i=1}^N \max\{p_i(0),p_i(1)\} \right) \notag\\
    & = 1/2 + \frac{1}{N}\left(\sum_{i=1}^N \min\{p_i(0),p_i(1)\}\right)
\end{align}
For the bit string that are not generated by $\mathsf{Embed}$ with the same key $sk$, the randomness $r$ is independent from the bits. According to the workflow of \textsf{Detect-1bit}, the conditional expectation of $X$ is 
\begin{align}
    & \mathbb{E}\left[X \mid b \xleftarrow{\$} \textsf{Unif}\{0,1\}^N \right] \notag\\
    =&\frac{1}{N}\sum_{i = 1}^N 0 \cdot \left(\mathbf{Pr}[r_i > 1/2] \cdot \mathbf{Pr}[b_i=1] + \mathbf{Pr}[r_i \leq 1/2] \cdot \mathbf{Pr}[b_i=0]\right) \notag\\
    &+ 1 \cdot \left(\mathbf{Pr}[r_i > 1/2] \cdot \mathbf{Pr}[b_i=0] + \mathbf{Pr}[r_i \leq 1/2] \cdot \mathbf{Pr}[b_i=1]\right)\notag\\
    =&1/2.
\end{align}

Obviously, the expectations of $X$ when embedding a watermark signal $m = 0$ and $m = 1$ are symmetric about $1/2$, which is the expectation of $X$ when computing on an unwatermarked bit string. And there is a gap $gap = \frac{1}{N}\left(\sum_{i=1}^N \min\{p_i(0),p_i(1)\} \right)$. As long as the \textsf{Predictor} does not always produce distributions like $(p(0)=0,p(1)=1)$, $gap > 0$ and 
these 3 conditional expectations have a noticeable gap between them. By applying Hoeffding's Inequality (Lemma \ref{Ho}), under the condition of $m = \bot$ and $t\in(0,1/2)$, the probability that sample mean $X$ deviates from its expectation up to $t$ is
\begin{align}
    \mathbf{Pr}\left[ X \geq \mathbb{E}[X\mid b \xleftarrow{\$} \textsf{Unif}\{0,1\}^N] + t\right] \leq \exp\left(-2N\cdot t^2\right),
\end{align}
which is negligible in $N$ if $t$ is a positive constant. If the $\mathsf{Predictor}$ does not always produce extreme distributions, $gap$ is a positive constant. It means that distinguishing the 3 different watermark signal $0,1$ and $\bot$ with an overwhelming probability is possible. 

Then we consider how to design the main algorithms $\mathsf{Embed}_{sk}$ and $\mathsf{Detect}_{sk}$.

\textbf{Construction of $\mathsf{Embed}_{sk}$ and $\mathsf{Detect}_{sk}$}.
To share the random numbers $r_i$, we let the $\mathsf{Embed}_{sk}$ and $\mathsf{Detect}_{sk}$ run a \textsf{PRF} with a key $sk$ generated by \textsf{KeyGen}.

As the algorithm $\mathsf{Detect}_{sk}$ does not know the distributions $(p_i(0), p_i(1))$, it cannot compute $\mathbb{E}\left[X \mid \mathsf{Embed}_{sk}(z,m=0)\right]$ or $\mathbb{E}\left[X \mid \mathsf{Embed}_{sk}(z,m=1)\right]$. $\mathsf{Detect}_{sk}$ cannot directly determine whether $m= 1$ or $m=0$. However, $\mathsf{Detect}_{sk}$ is able to compute the sample mean $X$ and compare it with $1/2$ to distinguish between $m= 0$ and $m=1$. In the end, $\mathsf{Detect}_{sk}(\mathbf{b'})$ works as follows:
\begin{itemize}
\item If $\exp\left(-2N\cdot (X-1/2)^2\right) \geq \mathsf{negl}(\lambda)$, output $m=\bot$.
    \item If $X>1/2$ and $\exp\left(-2N\cdot (X-1/2)^2\right) < \mathsf{negl}(\lambda)$, output $m=0$ and $block = \mathbf{b'}_{1:N}$.
    \item If $X<1/2$ and $\exp\left(-2N\cdot (X-1/2)^2\right) < \mathsf{negl}(\lambda)$, output $m=1$ and $block = \mathbf{b'}_{1:N}$.
\end{itemize}
$\exp\left(-2N\cdot (X-1/2)^2\right) < \mathsf{negl}(\lambda)$ represents the probability that $\mathsf{Detect}_{sk}$ incorrectly outputs a watermark signal $m=0$ or $m=1$ under the condition of $m = \bot$. 

Based on the above analysis, algorithm $\mathsf{Embed}_{sk}$ should continuously execute $\mathsf{Wat-Sampler}$ with the same valid watermark signal $m \in \{0,1\}$ as input. It can patiently generate many watermarked bits until the probability bound  $\exp\left(-2N\cdot t^2\right)$ is less than $\mathsf{negl}(\lambda)$.
As the $\mathsf{Embed}$ always patiently generates sufficient bits to ensure that probability is negligible in $\lambda$, without any modification on the watermarked bits, the $\mathsf{Detect}_{sk}$ can always output the correct watermark signal under the condition of $m = 0$ and $m=1$.

In summary, our multi-bit watermarking scheme $\mathcal{W} = (\mathsf{KeyGen},$ $ \mathsf{Embed}, \mathsf{Detect})$ works as follows:
\begin{itemize}
    \item $\mathsf{KeyGen}$ generates a secret watermark key $sk$.
    \item $\mathsf{Embed}$ substitutes the \textsf{Sampler} of \textsf{Model} for the \textsf{Wat-Sampler}, and runs \textsf{Model} with key $sk$, watermark signal $m \in \{0,1\}$ and some input $z$ to generate $N$ watermarked bits that satisfy $\exp\left(-2N\cdot (X-1/2)^2\right) < \mathsf{negl}(\lambda)$.
    \item $\mathsf{Detect}$ takes the $N$ bits as input and runs \textsf{Detect-1bit} to compute the watermark signal $X_i$ from each bit. Then it computes $X = \frac{1}{N}\sum_{i=1}^N X_i$ and $\exp\left(-2N\cdot (X-1/2)^2\right)$. After some comparison, it outputs a watermark signal $m\in\{0,1,\bot\}$. If $m\neq\bot$, it outputs the substring that is enough to detect the watermark signal.
\end{itemize}
This construction relies on 2 assumptions: (1) the security of \textsf{PRF} and (2) the \textsf{Predictor} does not always output distributions that has no entropy.

Details of $\mathsf{Embed}_{sk}$ and $\mathsf{Detect}_{sk}$ are shown in Alg. \ref{emb} and \ref{det}. In the following part of paper, we call the watermarked output of $\mathsf{Embed}_{sk}(m,z)$ as a watermark block that correctly contains a watermark signal $m$. In most of cases, the algorithm $\mathsf{Embed}_{sk}$ may be executed several times to generate several watermark blocks with a series of watermark signals. And the algorithm $\mathsf{Detect}_{sk}$ can also be executed multiple times to output a series of watermark signals. 



\begin{algorithm}[ht]
\caption{$\textsf{Wat-Sampler}((p(0),p(1)), m, r)$}
\label{watsamp}
    \renewcommand{\algorithmicrequire}{\textbf{Input:}}
    \renewcommand{\algorithmicensure}{\textbf{Output:}}
\begin{algorithmic}[1]
\REQUIRE distribution of next bit $(p(0),p(1))$, watermark signal $m$, randomness $r$.
\ENSURE a watermarked bit $b$.
\RETURN $b = \mathds{1}(r < p(b)) \oplus m$
\end{algorithmic}
\end{algorithm}

\begin{algorithm}[ht]
\caption{$\textsf{Detect-1bit}(b, r)$}
\label{detect1bit}
    \renewcommand{\algorithmicrequire}{\textbf{Input:}}
    \renewcommand{\algorithmicensure}{\textbf{Output:}}
\begin{algorithmic}[1]
\REQUIRE a bit $b$, randomness $r$.
\ENSURE a watermarked signal $m$.
\RETURN $m = \mathds{1}(r < 1/2) \oplus b$
\end{algorithmic}
\end{algorithm}

\begin{algorithm}[ht]
\caption{$\mathsf{Embed}_{sk}( m, z)$}
\label{emb}
    \renewcommand{\algorithmicrequire}{\textbf{Input:}}
    \renewcommand{\algorithmicensure}{\textbf{Output:}}
\begin{algorithmic}[1]
\REQUIRE watermark secret key $sk$, watermark signal $m \in \{0,1\}$, auxiliary input $z \in \{0,1\}^*$ for \textsf{Predictor}.
\ENSURE watermarked bit string $b$.
\STATE $b \longleftarrow \emptyset$
\REPEAT 
\STATE $(p(0),p(1)) \longleftarrow \mathsf{Predictor}(z)$
\STATE $r_i \longleftarrow \textsf{PRG}(sk)$
\STATE $b_i \longleftarrow \textsf{Wat-Sampler} \left((p(0),p(1)), m ,r_i\right)$
\STATE $b \longleftarrow b||b_i$
\STATE $z \longleftarrow z||b_i$
\STATE $X_i \longleftarrow \textsf{Detect-1bit}(b_i, r_i)$
\STATE $X \longleftarrow \frac{1}{i}\sum_{n = 1}^i X_i$
\UNTIL{$\exp\left(- \frac{i\cdot(X-1/2)^2}{2}\right) < \textsf{negl}(\lambda)$}
\RETURN $b$
\end{algorithmic}
\end{algorithm}

\begin{algorithm}[ht]
\caption{$\mathsf{Detect}_{sk}( b)$}
\label{det}
    \renewcommand{\algorithmicrequire}{\textbf{Input:}}
    \renewcommand{\algorithmicensure}{\textbf{Output:}}
\begin{algorithmic}[1]
\REQUIRE watermark secret key $sk$, bit string $b$.
\ENSURE watermark signal $m \in \{0,1,\bot\}$, watermark block $block$.
\STATE $m\longleftarrow \bot$
\FOR{$i \in \{1,2,\cdots, |b|\}$}
\STATE $r_i \longleftarrow \textsf{PRG}(sk)$
\STATE $X_i \longleftarrow \textsf{Detect-1bit}(b_i, r_i)$
\STATE $X \longleftarrow \frac{1}{i}\sum_{n = 1}^i X_i$
\IF{$\exp\left(- \frac{i\cdot(X-1/2)^2}{2}\right) < \textsf{negl}(\lambda)$}
    \IF {$X > 1/2$}
    \STATE $m \longleftarrow 0$
    \ELSE
    \STATE $m \longleftarrow 1$
    \ENDIF
    \STATE $hpos \longleftarrow i$
    \STATE \textbf{break}
\ENDIF
\ENDFOR
\STATE $block \longleftarrow b_{1:hpos}$
\RETURN $m, block$
\end{algorithmic}
\end{algorithm}

\subsection{Properties of Basic Construction}
In this section, we prove that our construction is correct, computational indistinguishable from the output of \textsf{Model}, and robust within a certain range.

The mistakes that \textsf{Detect} may make can be divided into 3 parts: 
\begin{itemize}
    \item (1) \textsf{Detect} outputs a watermark signal $m\in\{0,1\}$ from an unwatermarked bit string,
    \item (2) \textsf{Detect} outputs $m=\bot$ from a watermarked bit string,
    \item (3) \textsf{Detect} outputs $m=1$ (or $m=0$) from a bit string contains watermark signal $m = 0$ (or $m=1$).
\end{itemize} 
Therefore, we define the correctness as follows.
\begin{definition}[Correctness]
    A single-bit watermark scheme $\mathcal{W} = (\mathsf{KeyGen}, \mathsf{Embed}, \mathsf{Detect})$ is correct if for all possible bit string $b$ with length $|b| < \textsf{poly}(\lambda)$, 
    \begin{align}
        \mathbf{Pr}\left[ \mathsf{Detect}_{sk}(b) \neq m \mid b= \mathsf{Embed}_{sk}(z,m), m\in\{0,1\} \right] < \mathsf{negl}(\lambda),
    \end{align}
    and 
    \begin{align}
        \mathbf{Pr}\left[ \mathsf{Detect}_{sk}(b) \neq \bot \mid b \xleftarrow{\$} \mathsf{Unif}\{0,1\}^{\mathsf{poly}(\lambda)} \right] < \mathsf{negl}(\lambda),
    \end{align}
\end{definition}


\begin{claim}
    The proposed single-bit watermark scheme $\mathcal{W} = (\mathsf{KeyGen},$ $ \mathsf{Embed}, \mathsf{Detect})$ is correct.
\end{claim}
\textit{Proof.}
(1) In the above subsection, the probability that $\mathsf{Detect}_{sk}$ incorrectly outputs a watermark signal $m=0$ or $m=1$ from a single unwatermarked bit string is set less than a negligible function of $\lambda$. When making at most $\mathsf{poly}(\lambda)$ different queries to $\mathsf{Detect}_{sk}$ with a fixed $sk$, the probability of $\mathsf{Detect}_{sk}$ output a watermark signal $m = \bot$ is $\left( 1 - \mathsf{negl}(\lambda) \right)^{\mathsf{poly}(\lambda)}$, which is an overwhelming probability. This result shows that $\mathsf{Detect}_{sk}$ is still correct when processing $\mathsf{poly}(\lambda)$ different queries.
(2) For those watermarked bits generated by $\mathsf{Embed}_{sk}$, they are pre-detected by $\mathsf{Embed}_{sk}$ itself and the $\mathsf{Detect}_{sk}$ can always correctly detect them. (3) As for the condition that $\mathsf{Detect}_{sk}$ outputs a reverse watermark signal, its probability is less than the probability that $\mathsf{Detect}_{sk}$ outputs a watermark signal $m \in \{0,1\}$ from an unwatermarked bit string, which is negligible in $\lambda$.

In the end, the probability that $\mathsf{Detect}_{sk}$ makes mistakes is negligible in $\lambda$, which proves that the proposed single-bit watermark scheme is correct.

\begin{figure}[htbp]
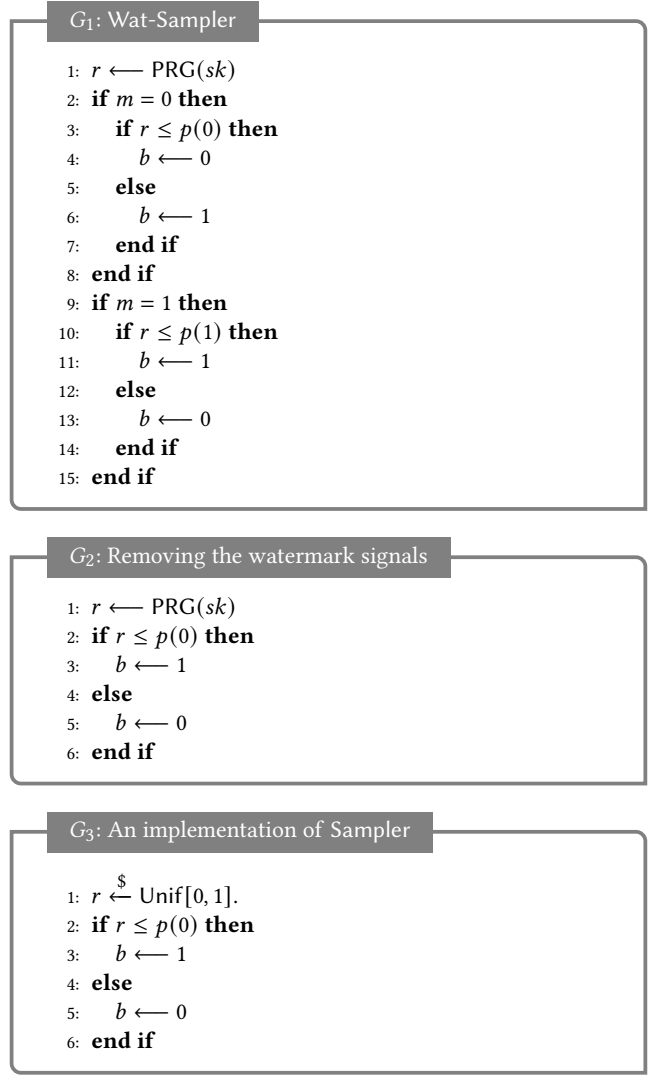

    \centering
    \begin{theobox}{$G_1$: \textsf{Wat-Sampler}}
    \begin{algorithmic}[1]
        \STATE $r\longleftarrow \textsf{PRG}(sk)$
        \IF{$m = 0$}
            \IF {$r \leq p(0) $}
                \STATE $b \longleftarrow 0$
            \ELSE
                \STATE $b \longleftarrow 1$
            \ENDIF
        \ENDIF
        \IF{$m = 1$}
            \IF {$r \leq p(1)$}
                \STATE $b \longleftarrow 1$
            \ELSE
                \STATE $b \longleftarrow 0$
            \ENDIF
        \ENDIF
    \end{algorithmic}
    \end{theobox}
    \begin{theobox}{$G_2$: Removing the watermark signals}
    \begin{algorithmic}[1]
    \STATE $r\longleftarrow \textsf{PRG}(sk)$
            \IF {$r \leq p(0)$}
                \STATE $b \longleftarrow 1$
            \ELSE
                \STATE $b \longleftarrow 0$
            \ENDIF
    \end{algorithmic}
    \end{theobox}
    \begin{theobox}{$G_3$: An implementation of \textsf{Sampler}}
    \begin{algorithmic}[1]
    \STATE $r \xleftarrow{\$} \mathsf{Unif}[0,1]$.
            \IF {$r \leq p(0)$}
                \STATE $b \longleftarrow 1$
            \ELSE
                \STATE $b \longleftarrow 0$
            \ENDIF
    \end{algorithmic}
    \end{theobox}
    \caption{Games used in the proof of computational indistinguishability.}
    \label{games}
\end{figure}

\begin{definition}[Undetectability] A watermark scheme $\mathcal{W} = (\mathsf{KeyGen}, \mathsf{Embed}, \mathsf{Detect})$ based on $\mathsf{Model}$ is undetectable if for all polynomial time adversaries $\mathcal{A}$,
\begin{align}
    \left|\mathbf{Pr}\left[\mathcal{A}^{\mathsf{Embed}_{sk}(\cdot,\cdot)}(1^\lambda)  = 1\right] - \mathbf{Pr}\left[\mathcal{A}^{\mathsf{Model}(\cdot)} (1^\lambda)  = 1\right] \right| < \mathsf{negl}(\lambda)
\end{align}
\end{definition}
\begin{claim}
    The proposed single-bit watermark scheme $\mathcal{W} = (\mathsf{KeyGen},$ $ \mathsf{Embed}, \mathsf{Detect})$ is undetectable.
\end{claim}
\textit{Proof}. Consider 3 games presented in Fig. \ref{games}. 
$G_1 = G_2$: $G_1$ is \textsf{Wat-Sampler} and $G_2$ is a pseudorandom sampler that does not take a watermark signal as an additional input. The probability that $G_1$ outputs 0 is 
\begin{align}
    & \mathbf{Pr}\left[ G_1 \text{outputs 0} \right] \notag \\
    = & \mathbf{Pr}\left[ G_1 \text{outputs 0}  \mid  m = 0\right] \mathbf{Pr}\left[ m = 0\right] \notag \\
   & + \mathbf{Pr}\left[ G_1 \text{outputs 0}  \mid  m = 1\right] \mathbf{Pr}\left[ m = 1\right]\notag \\
    = & \mathbf{Pr}\left[ r \leq p(0)\right]\mathbf{Pr}\left[ m = 0\right] + \mathbf{Pr}\left[ r > p(1)\right]\mathbf{Pr}\left[ m = 1\right]\notag \\
    = & \mathbf{Pr}\left[ r \leq p(0)\right]\left(\mathbf{Pr}\left[ m = 0\right] + \mathbf{Pr}\left[ m = 1\right]\right)\notag \\
    = & \mathbf{Pr}\left[ r \leq p(0)\right].
\end{align}
Obviously, the probability that $G_2$ outputs 0 is also $\mathbf{Pr}\left[ r \leq p(0)\right]$. Therefore, the output distributions of $G_1$ and $G_2$ are identical.

$G_2 \approx G_3$: $G_2$ uses a pseudorandom distribution to perform inverse transform sampling, while $G_3$ uses a real random distribution $\textsf{Unif}[0,1]$. Assuming that there exists an polynomial time adversary $\mathcal{A}$, which has a noticeable probability of distinguishing between $G_2$ and $G_3$. It is possible construct another polynomial time adversary $\mathcal{A}'$ to distinguish pseudorandom numbers and real random numbers. For example, adversary $\mathcal{A}'$ queries \textsf{PRG} and a real random oracle $\mathcal{O}$ to get a pair of random numbers. Though $\mathcal{A}'$ cannot directly distinguish them, it can run $G_2$ with these 2 numbers. If the used number is real random, it actually runs $G_3$. Now $\mathcal{A}'$ gets a pair of output from $G_2$ and $G_3$, and it runs $\mathcal{A}$ to distinguish between them. As $\mathcal{A}$ has a noticeable probability of distinguishing $G_2$ and $G_3$, $\mathcal{A}'$ can distinguish between pseudorandom numbers and real random numbers with a noticeable probability. However, according to the definition of \textsf{PRG}, there does not exist a polynomial time adversary that is able to distinguish between the output of \textsf{PRG} and real random numbers. So the adversary $\mathcal{A}$ does not exist, and the output distributions of $G_2$ and $G_3$ are computationally indistinguishable. 

In summary, $G_1 = G_2 \approx G_3$. The single-bit watermark scheme $\mathcal{W} = (\mathsf{KeyGen}, \mathsf{Embed}, \mathsf{Detect})$ is undetectable.

\begin{definition} [Hamming Ball]
    For a bit string $b$ with length $|b|$, the Hamming ball of $b$ with radius $\gamma$ is 
    \begin{align}
        B(b,\gamma ) = \{b'\in \{0,1\}^{|b|}: \mathsf{wt}(b\oplus b') \leq \gamma \}.
    \end{align}
\end{definition}
\begin{definition} [$(\gamma,\epsilon)$-Robustness of Single Watermark Output, against Substitution]
    A bit string $b \in \{ \mathsf{Embed}_{sk}(z,m): m\in\{0,1\}, z \in \{0,1\}^* \}$ is $(\gamma,\epsilon)$-robust against substitution if any bit string $b' \in B(b,\gamma)$, 
    \begin{align}
        \mathbf{Pr}\left[\mathsf{Detect}_{sk}(b') \neq m \mid b = \mathsf{Embed}_{sk}(z,m)\right] < \left[\mathsf{negl}(\lambda)\right]^\epsilon.
    \end{align}
\end{definition}

Intuitively, $\gamma$ indicates the level of robustness, and $\epsilon$ represents the loss of false positive rate if such robustness is guaranteed.

\begin{definition}[$(\gamma,\epsilon)$-Robustness of a Watermark Scheme, against Substitution]
    A watermark scheme $\mathcal{W} = (\mathsf{KeyGen},$ $ \mathsf{Embed}, \mathsf{Detect})$ is $(\gamma,\epsilon)$-Robustness against substitution, if for any watermarked bit string $b \in \{0,1\}^{\mathsf{poly}(\lambda)}$ that contains watermark signal $m$, 
    \begin{align}
        \underset{\gamma'}{\inf} \left\{ \mathbf{Pr}\left[\mathsf{Detect}(b') \neq m \mid b' \in B(b,\gamma') \right] <\left[\mathsf{negl}(\lambda)\right]^\epsilon \right\} = \gamma.
    \end{align}
\end{definition}

\begin{claim}
    $b = \mathsf{Embed}_{sk}(z,m)$ is $\left(\sqrt{\frac{-|b| \ln(\mathsf{negl}(\lambda))}{8}}, \frac{1}{4}\right)$-robust against substitution.
\end{claim}
\textit{Proof}. From the workflow of $\textsf{Detect-1bit}$, if the received bit $b_i$ is flipped, its output $X'_i$ is also flipped. So for the a bit string $b$ embedded with watermark signal $1$, it satisfies that $\exp\left(-2|b|\cdot (X-1/2)^2\right) < \mathsf{negl}(\lambda)$ and $X > 1/2$. We transform the above condition into the following inequality:
\begin{align}
    X > \frac{1}{2} + \sqrt{\frac{-\ln\left(\mathsf{negl(\lambda)}\right)}{2|b|}}.
\end{align}
For any bit string $b' \in B(b,\gamma)$ , the $\mathsf{Detect}$ may compute a distorted $X'$. Because $ \mathsf{wt}(b' \oplus b) \leq \gamma$, $X - X' \leq \frac{\gamma}{|b|}$. Meanwhile $X'$ should be still larger than $1/2$, otherwise the $\mathsf{Detect}$ will not output a correctwatermark signal. We write this condition as:
\begin{align}
    X' > \frac{1}{2} + \sqrt{\frac{-\ln\left(\mathsf{negl(\lambda)}\right)}{2|b|}} - \frac{\gamma}{|b|}
\end{align}

According to the definition of robustness, we compute the decision bound of $X'$ as:
\begin{align}
    \exp\left(-2|b|\cdot (X'-1/2)^2\right) > \exp\left(-2|b|\cdot (\sqrt{\frac{-\ln\left(\mathsf{negl(\lambda)}\right)}{2|b|}} - \frac{\gamma}{|b|})^2\right).
\end{align}

Let $\gamma$ be $\sqrt{\frac{-|b| \ln(\mathsf{negl}(\lambda))}{8}}$, we have
\begin{align}
    \exp\left(-2|b|\cdot (X'-1/2)^2\right) > \exp\left(-\frac{1}{4}\ln\left(\mathsf{negl}(\lambda)\right)\right) = \left[\mathsf{negl}(\lambda)\right]^{\frac{1}{4}}
\end{align}
Therefore, the bit string $b = \mathsf{Embed}_{sk}(z,m)$ is $\left(\sqrt{\frac{-|b| \ln(\mathsf{negl}(\lambda))}{8}}, \frac{1}{4}\right)$-robust against substitution.  \qed

\begin{claim}
    The proposed single-bit watermark scheme $\mathcal{W} = (\mathsf{KeyGen},$ $ \mathsf{Embed}, \mathsf{Detect})$ is 
    $\left(-\frac{1}{2}\ln(\mathsf{negl}(\lambda)),\frac{1}{4}\right)$-robust against substitution.
\end{claim}
\textit{Proof.} As for any watermarked bit string $b$, it is $\left(\sqrt{\frac{-|b| \ln(\mathsf{negl}(\lambda))}{8}}, \frac{1}{4}\right)$-robust against substitution. There exists a requirement for the minimum length of a watermarked bit string, because $\mathsf{negl}(\lambda) > \exp\left(-2|b|\cdot (X-1/2)^2\right) > \exp\left(-|b|/2\right)$. This requirement indicates that $|b| > -2\ln(\mathsf{negl}(\lambda))$. Let $\gamma$ be $\sqrt{\frac{-|b| \ln(\mathsf{negl}(\lambda))}{8}}$, we have 
\begin{align}
    \gamma = \sqrt{\frac{-|b| \ln(\mathsf{negl}(\lambda))}{8}} > -\frac{1}{2} \ln(\mathsf{negl}(\lambda)).
\end{align}
Therefore, the single-bit watermark scheme $\mathcal{W}$ is 
    $\left(-\frac{1}{2}\ln(\mathsf{negl}(\lambda)),\frac{1}{4}\right)$-robust against substitution. \qed


\vspace{0.5cm}


In conclusion, our basic construction maintains the properties of a traditional robust watermark, and its output maintains the distribution of the original \textsf{Model}. This construction naturally contains 3 states and 2 of them are symmetric. For the current popular zero-bit watermark methods \cite{kirchenbauer2023watermark,christ2024pseudorandom}, expanding them into a multi-bit version by using watermarked block as 0, unwatermarked block as 1 may affect its robustness, as those unwatermarked blocks cannot be detected. 

\subsection{An Simple Unforgeable Construction}
Our basic construction does not provide unforgeability, but it is feasible to generate multiple watermark blocks with certain watermark signals that represents a hash of the previous watermark blocks or prompt $z$. In the following part of the paper, we call the watermark blocks that contains a complete hash value of the previous watermark blocks as a watermark \textbf{link}, which has a fixed number of watermark blocks. And the entire construction, including the prompt $z$ and all watermark blocks (or links), is called a watermark \textbf{chain}. For simplicity, we set the length of $\mathsf{Hash}(\cdot)$ as the secure parameter $\lambda$.

\begin{definition}[Unforgeable Watermarking Scheme]
    An unforgeable watermarking scheme based on a model $\mathsf{Model}$ is a tuple of efficient algorithms $\mathcal{W}_{\mathsf{U}}$ = $(\mathsf{KeyGen}, \mathsf{UEmbed}, \mathsf{UDetect}, \mathsf{Verify})$ where:
    \begin{itemize}
    \item $\mathsf{KeyGen}(1^\lambda) \xrightarrow{} sk$ is a randomized algorithm that takes a security parameter $\lambda$ as input, and outputs a watermark key $sk$.
    \item $\mathsf{UEmbed}_{sk}(z) \xrightarrow{} b$ is a randomized algorithm that takes the prompt $z$ as input, and outputs a bit string $b$. 
    \item $\mathsf{UDetect}_{sk}(b') \xrightarrow{} (\{m_j\}_{j= 1}^*, \{block_{j}\}_{j= 1}^*)$ is a randomized algorithm that takes a bit string $b$ as input, and outputs a list of tuples $\{m_j, block_{j}\}_{j= 1}^*$, where $m_j$ represents the watermark signal, $block_{j}$ denotes the corresponding watermark block.
    \item $\mathsf{Verify}(z,\{m_j\}_{j= 1}^*, \{block_{j}\}_{j= 1}^*) \xrightarrow{} \{\mathsf{True},\mathsf{False}\}$ takes watermark signals and corresponding watermark blocks as input, and outputs $\mathsf{True}$ or $\mathsf{False}$.
\end{itemize}
\end{definition}

We construct a multi-bit watermarking scheme $\mathcal{W}_{\textsf{U}}$ = (\textsf{KeyGen}, \textsf{UEmbed}, \textsf{UDetect}, \textsf{Verify}), which works as follows:
\begin{itemize}
    \item $\mathsf{KeyGen}(1^\lambda) \xrightarrow{} sk$ generates a watermark key $sk$.
    \item $\mathsf{UEmbed}_{sk}(z) \xrightarrow{} b$ takes the prompt $z$ as input, and outputs a bit string $b$. First it let $prev$ be the prompt $z$. Then it computes \textsf{Hash}$(prev)$ and use each bit in \textsf{Hash}$(z)$ as watermark signal $m$ to run $\mathsf{Embed}_{sk}(m,z)$ to generate a series of watermark blocks, which is named as a watermark link. Then it updates the $prev$ to be the newest generated watermark link and repeats the above procedure until the \textsf{Wat-Model} naturally halts. In the end, this algorithm generates a chain of watermark links that each watermark link contains the hash of the preceding watermark link.
    \item $\mathsf{UDetect}_{sk}(b') \xrightarrow{} \{m_j, block_{j}\}_{j= 1}^*$ takes a bit string $b$ as input. It runs $\mathsf{Detect}_{sk}(b')$ on each substring $b_{i:|b|}$ for all $i \in \{1,2,\cdots,|b|\}$ to extract watermark signals. After it successfully extracts a watermark signal $m \in \{0,1\}$, it stores the $m$, and the corresponding watermark block. It outputs each watermark signal and the corresponding watermark block.
    \item $\mathsf{Verify}(z,\{m_j\}_{j= 1}^*, \{block_{j}\}_{j= 1}^*) \xrightarrow{} \{\mathsf{True},\mathsf{False}\}$ takes watermark signals and corresponding watermark blocks as input, and outputs $\mathsf{True}$ or $\mathsf{False}$. It first let $prev$ be the prompt $z$, and then computes $\mathsf{Hash}(prev)$. If $\mathsf{Hash}(prev) \neq m_{1:\lambda}$, it output $\mathsf{False}$. Otherwise, it let $prev$ be the first watermark link $||_{j = 1}^{\lambda}block_j$ and compare the $\mathsf{Hash}(prev)$ with the watermark signals embedded in the second watermark link. If except the last watermark link (which may not be complete because the \textsf{Wat-Model} halts) the other watermark link correctly contains the hash values, it outputs $\mathsf{True}$.
\end{itemize}

Details are shown in Alg. \ref{UEmbed} and \ref{UDetect}. In summary, $\mathsf{UEmbed}_{sk}$ uses hash values as watermark signals to generate watermarked content, and $\mathsf{UDetect}_{sk}$ scans the received bits from the beginning to the end to detect any watermark signal. It outputs watermark signals and the corresponding watermark blocks. $\mathsf{Verify}$ checks whether each watermark link are the successor of the previous watermark link or the prompt. 

Analyzing the output of $\mathsf{UDetect}_{sk}$ will provide more information about the condition of the input. If the $\mathsf{UDetect}_{sk}$ outputs an empty set, we confirm that the input are unwatermarked. If $\mathsf{UDetect}_{sk}$ outputs several watermark signals, but the watermark blocks does not cover all of the input bits, it means that the input has been severely modified. If $\mathsf{UDetect}_{sk}$ outputs lots of watermark signals, and the watermark blocks really cover all of the input bits, some minor modifications may have been made. In this condition, checking the hash values is necessary to confirm whether the input has been modified. 

\begin{algorithm}[th]
\caption{$\mathsf{UEmbed}_{sk}(z)$}
\label{UEmbed}
    \renewcommand{\algorithmicrequire}{\textbf{Input:}}
    \renewcommand{\algorithmicensure}{\textbf{Output:}}
\begin{algorithmic}[1]
\REQUIRE watermark secret key $sk$, auxiliary input $z$.
\ENSURE watermarked bits $b \in \{0,1\}^*$.
\STATE $prev \longleftarrow z$

\REPEAT
\STATE $h \longleftarrow \mathsf{Hash}(prev)$
\STATE $prev \longleftarrow \emptyset$
\FOR{$i \in \{1,2,\cdots,|h|\}$}
\STATE $block_i \longleftarrow \textsf{Embed}_{sk}(h_i, z)$
\STATE $z \longleftarrow z || block_i$
\STATE $prev \longleftarrow prev || block_i$
\ENDFOR
\STATE $b \longleftarrow b || prev$
\UNTIL{\textsf{Wat-Model} halts.}
\RETURN $b$
\end{algorithmic}
\end{algorithm}

\begin{algorithm}[th]
\caption{$\mathsf{UDetect}_{sk}(b)$}
\label{UDetect}
    \renewcommand{\algorithmicrequire}{\textbf{Input:}}
    \renewcommand{\algorithmicensure}{\textbf{Output:}}
\begin{algorithmic}[1]
\REQUIRE watermark secret key $sk$, bits $b$.
\ENSURE watermark signals and corresponding watermark blocks: $\{m_j\}_{j= 1}^*, \{block_{j}\}_{j= 1}^*$.
\STATE $cnt \longleftarrow 0$
\FOR{$i \in \{1,2,\cdots, |b|\}$}
\STATE $m', block' \longleftarrow \textsf{Detect}_{sk}(b_{i:|b|})$
\IF{$m_i \neq \bot$}
    \STATE $cnt \longleftarrow cnt + 1$
    \STATE $m_{cnt} \longleftarrow m'$
    \STATE $block_{cnt} \longleftarrow block'$
\ENDIF
\ENDFOR
\RETURN $\{m_j\}_{j= 1}^{cnt}, \{block_{j}\}_{j= 1}^{cnt}$
\end{algorithmic}
\end{algorithm}

\begin{algorithm}[th]
\caption{$\mathsf{Verify}(z,\{m_j\}_{j= 1}^k, \{block_{j}\}_{j= 1}^k)$}
\label{UDetect}
    \renewcommand{\algorithmicrequire}{\textbf{Input:}}
    \renewcommand{\algorithmicensure}{\textbf{Output:}}
\begin{algorithmic}[1]
\REQUIRE prompt $z$, watermark signals $\{m_j\}_{j= 1}^k$, watermark blocks $\{block_{j}\}_{j= 1}^k$.
\ENSURE \textsf{True} or \textsf{False}.

\STATE $prev \longleftarrow z$
\STATE $cnt \longleftarrow 1$
\STATE $flag \longleftarrow \mathsf{True}$
\WHILE{$cnt + \lambda \leq k$}
\STATE $hash \longleftarrow \mathsf{Hash}(prev)$
\IF{$hash \neq ||_{j=cnt}^{cnt + \lambda}m_j$}
    \STATE $flag \longleftarrow \mathsf{False}$
\ENDIF

\STATE $prev \longleftarrow ||_{j=cnt}^{cnt + \lambda}block_j$
\STATE $cnt \longleftarrow cnt + \lambda$
\ENDWHILE
\RETURN $flag$
\end{algorithmic}
\end{algorithm}

\subsection{Properties of Unforgeable Construction}

In this subsection, we first define the robustness-exploiting forgery attack and prompt-misattribution forgery attack, and then we prove that the proposed multi-bit construction is prefix-unforgeable.

\begin{definition}[Robustness-Exploiting Forgery Game]
    A polynomial time adversary $\mathcal{A}$ works as follows:
    \begin{itemize}
        \item $\mathcal{A}$ queries the $\mathsf{UEmbed}_{sk}$ with a prompt $z$, obtaining a watermarked output $b = \mathsf{UEmbed}_{sk}(z) $.
        \item $\mathcal{A}$ chooses any watermarked bit $b_i \in \mathsf{UEmbed}_{sk}(z)$, flips the bit $b_i$ and outputs the modified watermarked output $b'$.
        \item $\mathcal{A}$ claims the modified output $b'$ as a watermarked output without modification.
    \end{itemize}
    If $\mathsf{Verify}(z,\mathsf{UDetect}(b'))$ outputs $\mathsf{True}$, $\mathcal{A}$ wins. Otherwise, $\mathcal{A}$ loses.
\end{definition}

\begin{definition}[Prompt-Misattribution Forgery Game]
A polynomial time adversary $\mathcal{A}$ works as follows:
\begin{itemize}
    \item $\mathcal{A}$ queries the $\mathsf{UEmbed}_{sk}$ with a prompt $z$, obtaining a watermarked output $b = \mathsf{UEmbed}_{sk}(z) $.
    \item $\mathcal{A}$ chooses another prompt $z' \neq z$.
    \item $\mathcal{A}$ claims the watermarked output $b$ is the successor of $z'$.
\end{itemize}
    If $\mathsf{Verify}(z',\mathsf{UDetect}(b))$ outputs $\mathsf{True}$, $\mathcal{A}$ wins. Otherwise, $\mathcal{A}$ loses.
\end{definition}

\begin{definition}[$\mathsf{Prefix}$]
    For a complete watermark chain $z||\mathsf{UEmbed}_{sk}(z)$, we define the $\mathsf{Prefix}$ as a prefix of the watermark chain $z||\mathsf{UEmbed}_{sk}(z)$ which contains all of the complete watermark links except the last one.
\end{definition}

\begin{definition}[Prefix-unforgeability]
    A watermarking scheme is prefix-unforgeable, if for any polynomial time adversary $\mathcal{A}$, 
    \begin{align}
        \mathbf{Pr}&\big[\mathcal{A}^{\mathsf{UEmbed}_{sk}(\cdot)}(1^\lambda) \xrightarrow{} z'||b', \mathsf{Prefix}(b') \neq \mathsf{Prefix}(\mathsf{UEmbed}_{sk}(z')), \notag \\
        &\mathsf{Verify}(z',\mathsf{UDetect}_{sk}(b')) = \mathsf{True} \big] 
        < \mathsf{negl}(\lambda)
    \end{align}
\end{definition}

\begin{claim}
    The proposed multi-bit watermarking scheme $\mathcal{W}_{\mathsf{U}} = (\mathsf{KeyGen},$ $ \mathsf{UEmbed}, \mathsf{UDetect}, \mathsf{Verify})$ is prefix-unforgeable. 
\end{claim}
\textit{Proof.} For any bit string $b'$ and prompt $z'$, if $\mathsf{Prefix}(b') \neq \mathsf{Prefix}($ $\mathsf{UEmbed}_{sk}(z'))$, applying the $\mathsf{Hash}$ on some watermark link will derive a different hash value with an overwhelming probability. For a polynomial time adversary $\mathcal{A}$, it cannot find 2 bit strings that has the same hash value with a non-negligible probability, otherwise it will break the collision-resistance property of \textsf{Hash}. \qed

We have to limit the unforgeability only on the prefix of output, because the last watermark link does not have its successor. If we detect and verify the original output of $\mathsf{UEmbed}_{sk}$, we can only conclude that its prefix has not been modified. Other multi-bit watermark methods \cite{christ2024pseudorandom,fairoze2024publiclydetectablewatermarkinglanguagemodels,cohen2024watermarking} will face the same problem as the model-generated text has its end.

Such construction does not fully resist robustness-exploiting forgery attacks, but it can cover most of the forgery attempts. As for prompt-misattribution forgery attacks, as long as $\mathsf{Hash}(z') \neq \mathsf{Hash}(z)$ and the watermarked output contains the first watermark link, the adversary can only win with a negligible probability. It is worthy to mention that \cite{fairoze2024publiclydetectablewatermarkinglanguagemodels} cannot resist prompt-misattribution forgery attacks, as this construction only protects the generated content. We do not choose to use a complex signature scheme due to a realistic reason: the entropy of model output is limited. Embedding a single watermark signal may require 100 text bits or more to guarantee its correctness and robustness. For a complete signature scheme, its output consists of hundreds of bits which is not possible to embed in a single response of model.

The undetectability and correctness of the proposed multi-bit watermarking scheme inherits directly from the single-bit watermarking scheme. Thus we do not prove the following claim.

\begin{claim}
    The proposed multi-bit watermarking scheme $\mathcal{W}_{\mathsf{U}} = (\mathsf{KeyGen},$ $ \mathsf{UEmbed}, \mathsf{UDetect}, \mathsf{Verify})$ is undetectable if the single-bit watermarking scheme $\mathcal{W} = (\mathsf{KeyGen},$ $ \mathsf{Embed}, \mathsf{Detect})$ is undetectable.
\end{claim}

\begin{claim}
    The proposed multi-bit watermarking scheme $\mathcal{W}_{\mathsf{U}} = (\mathsf{KeyGen},$ $ \mathsf{UEmbed}, \mathsf{UDetect}, \mathsf{Verify})$ is correct if the single-bit watermarking scheme $\mathcal{W} = (\mathsf{KeyGen},$ $ \mathsf{Embed}, \mathsf{Detect})$ is correct.
\end{claim}

\section{Experiments \& Results}

\subsection{Implementation Details}

Current language models output a distribution which is defined on an alphabet (or a vocabulary list). 

We set the $\mathsf{negl}(\lambda)$ as $e^{-\lambda}$.

\newpage    
\bibliography{abbrev}
\bibliographystyle{plain}

\end{document}